\documentstyle[12pt,axodraw,psfig]{article}
\newcommand{\gsim}{ \mathop{}_{\textstyle \sim}^{\textstyle >} }
\newcommand{\lsim}{ \mathop{}_{\textstyle \sim}^{\textstyle <} }

\begin{document}

\begin{titlepage}

\begin{flushright}
UCB-PTH-01/14 \\
LBNL-47876 \\
\end{flushright}
\vskip 1.0cm

\begin{center}
{\Large \bf  Constrained Standard Model from Extra Dimension}
\vskip 1.0cm

\def\thefootnote{\fnsymbol{footnote}}
{\large 
Yasunori Nomura\footnote{Research Fellow,
  Miller Institute for Basic Research in Science.}
}
\vskip 0.5cm

{\it Department of Physics, \\ and \\
Theoretical Physics Group, Lawrence Berkeley National Laboratory,\\
University of California, Berkeley, CA 94720}

\vskip 1.0cm

\abstract{A five dimensional supersymmetric model is constructed which 
reduces to the one Higgs-doublet standard model at low energies.
The radiative correction to the Higgs potential is finite and calculable, 
allowing the Higgs mass prediction of $127 \pm 8$ GeV.
The physical reasons for this finiteness are discussed in detail.
The masses of the superparticles and the Kaluza-Klein excitations 
are also predicted.  The lightest superparticle is a top squark 
of mass $197 \pm 20$ GeV.}

\end{center}
\end{titlepage}

\def\thefootnote{\arabic{footnote}}
\setcounter{footnote}{0}

%%%%%%%%%%
%%%%%%%%%%      main part 
%%%%%%%%%%

\section{Introduction}

The standard model extremely well describes the physics down to a 
scale of $10^{-18}$ m.  To achieve the electroweak symmetry breaking 
(EWSB), it employs the Higgs sector which consists of one SU(2) doublet 
scalar field $h$ with the following Lagrangian,
\begin{equation}
  {\cal L}_{\rm Higgs} = - m_h^2 |h|^2 - \lambda_h |h|^4,
\label{eq:Higgs-Lag}
\end{equation}
where $m_h^2$ is assumed to be negative to trigger EWSB.  
This Higgs sector is the least explored sector of the standard model 
and has the following unpleasant features:
\begin{itemize} 
\item The radiative correction $\delta m_h^2$ to the Higgs mass-squared 
parameter is quadratically divergent, $\delta m_h^2 \propto -\Lambda^2$ 
($\Lambda$ is a cutoff scale), and completely dominated by 
unknown ultraviolet physics.  It means that we cannot reliably 
calculate the crucial quantity $m_h^2$ for EWSB, and thus the standard 
model does not provide the theory of EWSB.  Also, if the cutoff scale 
$\Lambda$ is large as in the conventional grand desert scenario, an 
extreme fine-tuning between the tree and loop level contributions 
is required to maintain the Higgs vacuum expectation value (VEV) 
at the weak scale.

\item The Lagrangian Eq.~(\ref{eq:Higgs-Lag}) contains two free parameters, 
$m_h^2$ and $\lambda_h$.  Since the observed Fermi constant determines 
only one linear combination, $\left\langle h \right\rangle^2 = 
|m_h^2/(2 \lambda_h)| \simeq (174~{\rm GeV})^2$, the other combination 
remains unfixed and we cannot predict the physical Higgs boson mass.
Although this is not a problem of the theory, it is somewhat unpleasant 
from the viewpoint of the predictivity of the model.
\end{itemize}

Supersymmetry (SUSY) elegantly solves the problem of quadratic 
divergence.  If we extend the standard model to be supersymmetric, 
the radiative correction to the Higgs mass cancels between the loops 
of bosons and fermions, so that the weak-scale Higgs mass becomes 
stable quantum mechanically.  However, since no degeneracy between 
bosons and fermions is seen in nature, SUSY must be a broken symmetry.
The size of the SUSY-breaking masses should be of the order of 
the weak scale to keep the Higgs mass at the weak scale.

This leads to the usual framework beyond the standard model, specifically 
to the minimal SUSY standard model (MSSM) or its extensions.
Since the quadratic divergence of $\delta m_h^2$ cancels in these theories, 
we can make the cutoff scale $\Lambda$ large, say the Planck scale, 
without introducing enormous fine-tuning.  Furthermore, the radiative 
correction $\delta m_h^2$ can be calculated and is negative due to 
top/stop loops \cite{radewsb}.  Therefore, we have a theory of EWSB.

Although the above framework is a promising candidate of the physics 
beyond the standard model, there are still several features which are not 
completely satisfactory.  They are, 
\begin{itemize} 
\item Since we need soft SUSY-breaking parameters of the order of the 
weak scale, we have to answer how SUSY is broken.  It is known that, 
to obtain a realistic spectrum, SUSY-breaking sector must be separated 
from the MSSM sector, and the breaking must be transmitted to the MSSM 
sector through some interactions \cite{Dimopoulos:1981zb}.
This makes the structure of the theory somewhat complicated.

\item Although the quadratic divergence in $\delta m_h^2$ cancels between 
bosonic and fermionic loops, there is still a residual logarithmic 
divergence.  This means that the contributions to $\delta m_h^2$ come 
from all energies between the weak and the cutoff scales, so that the 
scale of the EWSB physics is unclear in this type of models.

\item In supersymmetric models, quartic couplings of the Higgs fields 
are generically related to the standard model gauge couplings: 
$\lambda_h \sim (g^2+g'^2)/8$ where $g$ and $g'$ are the SU(2) and 
U(1) gauge couplings, respectively.  However, SUSY also requires two 
Higgs doublets to give both up- and down-type quark masses and to 
cancel gauge anomalies coming from Higgsinos.  This introduces extra 
free parameters such as the ratio of the VEVs of the two Higgs doublets.  
Thus, the physical Higgs boson mass is still not predicted (although 
there is an upperbound coming from the fact that $\lambda_h$'s are 
determined by SUSY).
\end{itemize}
The question then is whether these conclusions are really unavoidable 
in {\it any} SUSY extension of the standard model.

In fact, to reach the above conclusions, we have implicitly assumed 
that nature is four dimensional up to high energy scales 
such as the Planck scale.  More precisely, there is an energy interval 
where the physics is described by a four dimensional ${\cal N} = 1$ 
supersymmetric theory.  However, we now know that it does not necessarily 
have to be true.  For instance, if there are large extra dimensions 
in which only gravity propagates, the fundamental scale of nature 
can be a TeV scale \cite{Arkani-Hamed:1998rs}.  The observed 
weakness of the gravity then is explained by a wavefunction suppression 
of the graviton due to a large volume of the extra dimensions.

One might think that there is no need for SUSY in this case, since 
the divergence of the Higgs mass is cut off at the TeV scale.
However, without SUSY, the divergence is still quadratic, so that we 
even do not know the sign of $m_h^2$ in effective field theory.
Therefore, unless we embed the theory into a more fundamental theory 
such as string theory and calculate $\delta m_h^2$, the framework itself 
does not provide a theory of EWSB.  We still need SUSY to control the 
Higgs mass and have a theory of EWSB at the effective field theory level.

However, lowering the fundamental scale down to a TeV scale opens up 
a wide variety of possibilities as to how to implement SUSY to the 
standard model.  In particular, we can consider a TeV-sized extra 
dimension in which the standard-model gauge, quark and lepton multiplets 
propagate.  That is, the theory has a higher dimensional SUSY above 
the TeV scale.  (The TeV-scale extra dimension has been considered in 
Ref.~\cite{Antoniadis:1990ew}.)  Then, we find that the previous 
conclusions are no longer necessarily correct.  Specifically, 
we can construct a model \cite{Barbieri:2001vh} in which 
\begin{itemize} 
\item We do not need any hidden sector to break SUSY.  SUSY is broken 
by a compactification \cite{Scherk:1979ta}, and the masses for the 
superparticles are determined by the radius $R$ of the extra dimension 
\cite{SS-model,Barbieri:2001vh,Arkani-Hamed:2001mi}.

\item The radiative correction to the Higgs mass-squared parameter 
(the Higgs effective potential $V_h$) is completely finite and 
calculable in terms of the compactification radius $R$ 
\cite{Arkani-Hamed:2001mi,Barbieri:2001vh}.

\item The low-energy effective theory is precisely one Higgs-doublet 
standard model with the tree-level Higgs potential constrained as
\begin{equation}
  V_{h, {\rm tree}} = \frac{g^2+g'^2}{8} |h|^4.
\label{eq:Higgs-tree}
\end{equation}
\end{itemize}
Since the tree-level Higgs potential does not contain any free parameter 
and the radiative correction is calculated in terms of one parameter $R$, 
we can determine the compactification radius $R$ by requiring the Higgs 
VEV to be $\left\langle h \right\rangle \simeq 174~{\rm GeV}$.
It also determines the physical Higgs boson, superparticle and 
Kaluza-Klein (KK) excitation masses.  In the next section, we briefly 
review the structure of five dimensional SUSY and an orbifold 
compactification.  The explicit model of Ref.~\cite{Barbieri:2001vh} 
is explained in section \ref{sec:csm}.  Finally, section \ref{sec:concl} 
is devoted to our conclusions.

\section{Five Dimensional Supersymmetry}
\label{sec:review}

In this article, we consider supersymmetric models in five dimensions 
(5D) where all the standard-model fields propagate in the bulk.
The SUSY multiplets in 5D are larger than those in four dimensions (4D).
The standard model matter and Higgs fields are described by a set of 
hypermultiplets, and each hypermultiplet contains two complex scalars 
and two Weyl fermions.  For instance, the standard-model quark $q$ is 
accompanied by a squark $\tilde{q}$, a conjugate squark $\tilde{q}^c$ and 
a conjugate quark $q^c$.  Here, conjugated objects have conjugate 
transformations under the gauge group.  The standard model gauge bosons 
$A_\mu$ ($\mu = 0,1,2,3$) are contained in 5D vector supermultiplets 
$(A_M, \lambda, \lambda^c, \sigma)$ where $M = \mu, 5$.

\begin{figure}
\begin{center} 
\begin{picture}(350,150)(-10,-20)
  \Line(5,0)(320,0)
  \LongArrow(10,-10)(10,100)
  \Text(0,105)[b]{mass}
  \Text(0,0)[r]{$0$}
  \Line(8,20)(12,20)    \Text(6,20)[r]{$1/R$}
  \Line(8,40)(12,40)    \Text(6,40)[r]{$2/R$}
  \Line(8,60)(12,60)    \Text(6,60)[r]{$3/R$}
  \Line(8,80)(12,80)    \Text(6,80)[r]{$4/R$}
  \Text(60,100)[b]{$q, h, A_\mu$}
  \Line(40,0)(80,0)      \Vertex(60,0){3}
  \Line(40,20)(80,20)    \Vertex(50,20){3} \Vertex(70,20){3}
  \Line(40,40)(80,40)    \Vertex(50,40){3} \Vertex(70,40){3}
  \Line(40,60)(80,60)    \Vertex(50,60){3} \Vertex(70,60){3}
  \Line(40,80)(80,80)    \Vertex(50,80){3} \Vertex(70,80){3}
  \Text(130,100)[b]{$\tilde{q}, \tilde{h}, \lambda$}
  \Line(110,0)(150,0)      \Vertex(130,0){3}
  \Line(110,20)(150,20)    \Vertex(120,20){3} \Vertex(140,20){3}
  \Line(110,40)(150,40)    \Vertex(120,40){3} \Vertex(140,40){3}
  \Line(110,60)(150,60)    \Vertex(120,60){3} \Vertex(140,60){3}
  \Line(110,80)(150,80)    \Vertex(120,80){3} \Vertex(140,80){3}
  \Text(200,100)[b]{$\tilde{q}^c, \tilde{h}^c, \lambda^c$}
  \Line(180,0)(220,0)      \Vertex(200,0){3}
  \Line(180,20)(220,20)    \Vertex(190,20){3} \Vertex(210,20){3}
  \Line(180,40)(220,40)    \Vertex(190,40){3} \Vertex(210,40){3}
  \Line(180,60)(220,60)    \Vertex(190,60){3} \Vertex(210,60){3}
  \Line(180,80)(220,80)    \Vertex(190,80){3} \Vertex(210,80){3}
  \Text(270,100)[b]{$q^c, h^c, \sigma$}
  \Line(250,0)(290,0)      \Vertex(270,0){3}
  \Line(250,20)(290,20)    \Vertex(260,20){3} \Vertex(280,20){3}
  \Line(250,40)(290,40)    \Vertex(260,40){3} \Vertex(280,40){3}
  \Line(250,60)(290,60)    \Vertex(260,60){3} \Vertex(280,60){3}
  \Line(250,80)(290,80)    \Vertex(260,80){3} \Vertex(280,80){3}
\end{picture}
\caption{Tree-level KK mass spectrum of the matter, Higgs and gauge 
multiplets in $S^1$ compactification.}
\label{fig:S1}
\end{center}
\end{figure}
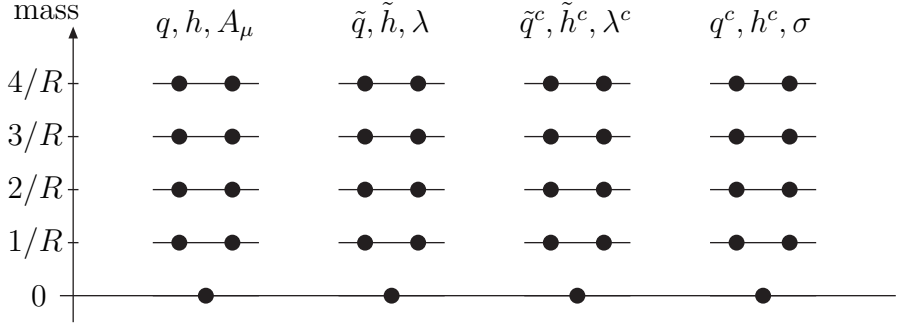

Upon compactifying the fifth dimension $y$ on the circle $S^1$, each 
5D field $\varphi(x^\mu, y)$ is decomposed into a tower of 4D fields 
$\varphi_n(x^\mu)$ as
\begin{equation}
  \varphi(x^\mu, y) = \sum_{n=-\infty}^{\infty} 
    \varphi_n(x^\mu)\, {\rm e}^{i n y/R},
\label{eq:KK-S1}
\end{equation}
where the mass of the 4D field $\varphi_n$ is given by $|n|/R$.
The resulting spectrum is given in Fig.~\ref{fig:S1}, where $q$ 
correctively represents the standard-model quark and lepton fields.
We find that there are many unwanted fields in the massless sector.
In particular, the theory is necessarily vector-like reflecting 
underlying 5D Lorentz invariance ($q$ is always accompanied by $q^c$), 
so that it cannot reproduce the standard model at low energies.
Therefore, the simple compactification on $S^1$ clearly does not work.

What we have to do then is to compactify the extra dimension on 
an orbifold.  The simplest orbifold called $S^1/Z_2$ is constructed by 
compactifying the extra dimension $y$ on a circle of radius $R$ 
and identifying points under the reflection $y \leftrightarrow -y$.
Then, the physical space is an interval $y[0, \pi R]$ and has two 
special points at $y=0$ and $\pi R$ called fixed points.
Under the reflection $y \leftrightarrow -y$, various 5D fields have 
definite transformation properties (parities), which are determined 
such that 5D Lagrangian is invariant under the parity transformation.
We assign positive and negative parities for unconjugated and conjugated 
objects, respectively, in matter and Higgs multiplets.
The transformations for the gauge multiplets are determined as 
$(+):A_\mu,\lambda,\,(-):\lambda^c,\sigma$.

\begin{figure}
\begin{center} 
\begin{picture}(350,160)(-10,-20)
  \Line(5,0)(320,0)
  \LongArrow(10,-10)(10,100)
  \Text(0,105)[b]{mass}
  \Text(0,0)[r]{$0$}
  \Line(8,20)(12,20)    \Text(6,20)[r]{$1/R$}
  \Line(8,40)(12,40)    \Text(6,40)[r]{$2/R$}
  \Line(8,60)(12,60)    \Text(6,60)[r]{$3/R$}
  \Line(8,80)(12,80)    \Text(6,80)[r]{$4/R$}
  \Text(60,115)[b]{$(+)$}
  \Text(60,100)[b]{$q, h, A_\mu$}
  \Line(40,0)(80,0)      \Vertex(60,0){3}
  \Line(40,20)(80,20)    \Vertex(60,20){3}
  \Line(40,40)(80,40)    \Vertex(60,40){3}
  \Line(40,60)(80,60)    \Vertex(60,60){3}
  \Line(40,80)(80,80)    \Vertex(60,80){3}
  \Text(130,115)[b]{$(+)$}
  \Text(130,100)[b]{$\tilde{q}, \tilde{h}, \lambda$}
  \Line(110,0)(150,0)      \Vertex(130,0){3}
  \Line(110,20)(150,20)    \Vertex(130,20){3}
  \Line(110,40)(150,40)    \Vertex(130,40){3}
  \Line(110,60)(150,60)    \Vertex(130,60){3}
  \Line(110,80)(150,80)    \Vertex(130,80){3}
  \Text(200,115)[b]{$(-)$}
  \Text(200,100)[b]{$\tilde{q}^c, \tilde{h}^c, \lambda^c$}
  \Line(180,0)(220,0)
  \Line(180,20)(220,20)    \Vertex(200,20){3}
  \Line(180,40)(220,40)    \Vertex(200,40){3}
  \Line(180,60)(220,60)    \Vertex(200,60){3}
  \Line(180,80)(220,80)    \Vertex(200,80){3}
  \Text(270,115)[b]{$(-)$}
  \Text(270,100)[b]{$q^c, h^c, \sigma$}
  \Line(250,0)(290,0)
  \Line(250,20)(290,20)    \Vertex(270,20){3}
  \Line(250,40)(290,40)    \Vertex(270,40){3}
  \Line(250,60)(290,60)    \Vertex(270,60){3}
  \Line(250,80)(290,80)    \Vertex(270,80){3}
\end{picture}
\caption{Tree-level KK mass spectrum of the matter, Higgs and gauge 
multiplets in $S^1/Z_2$ compactification.}
\label{fig:S1Z2}
\end{center}
\end{figure}
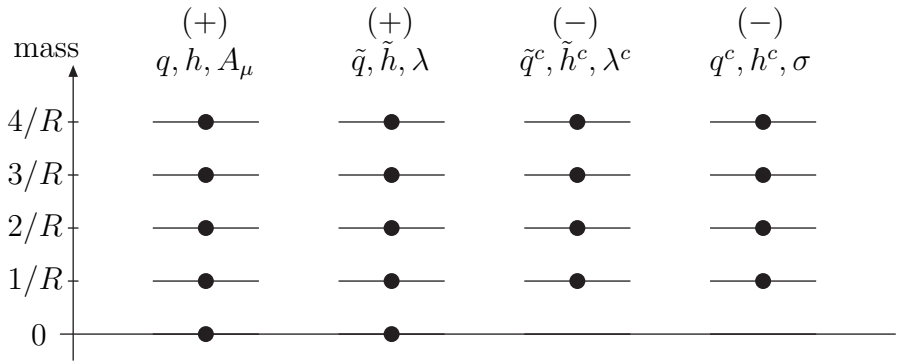

The decomposition of 5D fields now goes through as follows:
\begin{eqnarray}
  (+):&& \varphi(x^\mu, y) = \sum_{n=0}^{\infty} 
         \varphi_n(x^\mu)\, \cos \frac{n\, y}{R},
\\
  (-):&& \varphi(x^\mu, y) = \sum_{n=1}^{\infty} 
         \varphi_n(x^\mu)\, \sin \frac{n\, y}{R}.
\label{eq:KK-S1Z2}
\end{eqnarray}
The resulting mass spectrum is given in Fig.~\ref{fig:S1Z2}.
We find that the half of the states are projected out by the 
identification $y \leftrightarrow -y$.  Important point here is 
that, since $\sin 0$ is identically zero, the odd fields do not 
have zero modes.  Thus, the structure of the massless sector is 
reduced to that of 4D ${\cal N}=1$ SUSY and we obtain a chiral theory.
However, it is not sufficient, since 4D ${\cal N}=1$ SUSY is still 
remaining and we have massless superparticles.  We have to do 
something more to obtain a completely realistic theory.

\section{Constrained Standard Model}
\label{sec:csm}

We here propose to use the above orbifolding procedure twice to obtain 
a realistic theory \cite{Barbieri:2001vh}.  Then, the second orbifolding 
projects out further half of the states, which breaks SUSY completely, 
and the matter content of the zero-mode sector is reduced to that of 
the standard model.  Moreover, we can also introduce interactions which 
are precisely those of the one Higgs-doublet standard model, with the 
tree-level Higgs potential constrained as in Eq.~(\ref{eq:Higgs-tree}) 
due to the underlying SUSY structure.  Below, we will see how this 
works explicitly.

\subsection{$S^1/(Z_2 \times Z_2')$ orbifold}

\begin{figure}
\begin{center} 
\begin{picture}(450,100)(-260,-20)
% Figure 1
  \GOval(-218,17)(25,50)(-35){0.97}
  \GOval(-182,17)(25,50)(35){0.88}
  \Oval(-218,17)(25,50)(-35)
  \Text(-200,65)[b]{$q (+,+)$}
  \Text(-215,35)[r]{$\tilde{q} (+,-)$}
  \Text(-185,35)[l]{$\tilde{q}^{c\dagger} (-,+)$}
  \Text(-200,10)[t]{$q^{c\dagger} (-,-)$}
  \Line(-227,44)(-213,58) \DashLine(-173,44)(-187,58){2}
% Figure 2
  \GOval(-22,53)(25,50)(-35){0.97}
  \GOval(-22,17)(25,50)(35){0.88}
  \Oval(-22,53)(25,50)(-35)
  \Text(-40,65)[b]{$\tilde{h} (+,-)$}
  \Text(-55,35)[r]{$h (+,+)$}
  \Text(-25,35)[l]{$h^{c\dagger} (-,-)$}
  \Text(-40,10)[t]{$\tilde{h}^{c\dagger} (-,+)$}
  \Line(-67,44)(-53,58) \DashLine(-67,26)(-53,12){2}
% Figure 3
  \GOval(102,17)(25,50)(-35){0.97}
  \GOval(138,17)(25,50)(35){0.88}
  \Oval(102,17)(25,50)(-35)
  \Text(120,65)[b]{$A_{\mu} (+,+)$}
  \Text(105,35)[r]{$\lambda (+,-)$}
  \Text(135,35)[l]{$\lambda^c (-,+)$}
  \Text(120,10)[t]{$\Sigma (-,-)$}
  \Line(93,44)(107,58)  \DashLine(147,44)(133,58){2}
\end{picture}
\caption{Quantum numbers of the matter, Higgs and gauge 
multiplets under the two orbifoldings $y \leftrightarrow -y$ and 
$y' \leftrightarrow -y'$.  Here, $\Sigma = (\sigma + iA_5)/\sqrt{2}$.}
\label{fig:transf}
\end{center}
\end{figure}
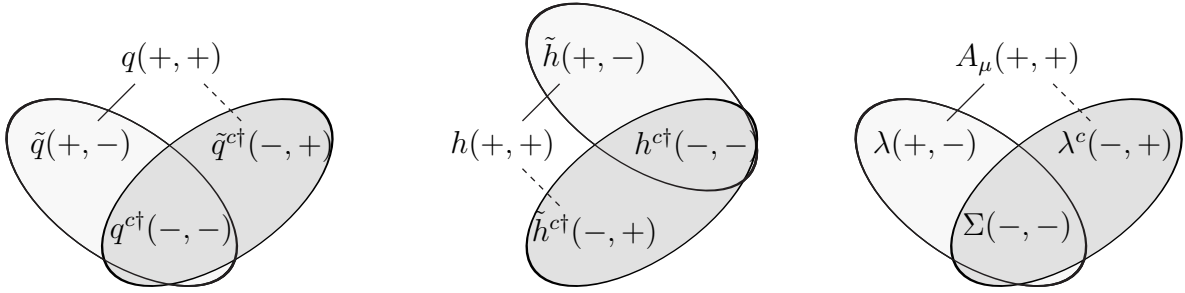

We consider a 5D SU(3)$\times$SU(2)$\times$U(1) gauge theory and 
introduce hypermultiplets corresponding to three generations of matter, 
$Q,U,D,L,E$ and a {\it single} Higgs field $H$.  The extra dimension is 
compactified on the $S^1/(Z_2 \times Z_2')$ orbifold, which is 
constructed by making two identifications, $Z_2: \; y \leftrightarrow -y$ 
and $Z_2': \; y' \leftrightarrow -y'$ ($y' = y - \pi R /2$), on a circle 
of radius $R$.  The physical space is now an interval $y[0, \pi R/2]$ 
which is a quarter of the circle.  The 5D fields have definite 
transformation properties (parities) under the $Z_2 \times Z_2'$, 
and they are decomposed as 
\begin{eqnarray}
  (+,+):&& \varphi(x^\mu, y) = \sum_{n=0}^{\infty} 
      \varphi_{2n}(x^\mu) \cos{2ny \over R},
\\
  (+,-):&& \varphi(x^\mu, y) = \sum_{n=0}^{\infty} 
      \varphi_{2n+1}(x^\mu) \cos{(2n+1)y \over R},
\\
  (-,+):&& \varphi(x^\mu, y) = \sum_{n=0}^{\infty} 
      \varphi_{2n+1}(x^\mu) \sin{(2n+1)y \over R},
\\
  (-,-):&& \varphi(x^\mu, y) = \sum_{n=0}^{\infty} 
      \varphi_{2n+2}(x^\mu) \sin{(2n+2)y \over R},
\label{eq:KK-S1Z2Z2}
\end{eqnarray}
according to their transformation properties under $(Z_2, Z_2')$.
The explicit parity assignment is given in Fig.~\ref{fig:transf}.
As we have seen in the previous section, the orbifolding by $Z_2$ 
projects out half of the states, which are inside the dark shaded ovals, 
from the zero-mode sector.  However, now we have the other orbifolding 
by $Z_2'$, which projects out different set of the states encircled 
by the light shaded ovals.  As a result, if we make both orbifoldings 
simultaneously, the matter content of the zero-mode sector 
($(+,+)$ fields) reduces precisely to that of the standard model.

\begin{figure}
\begin{center} 
\begin{picture}(350,160)(-10,-20)
  \Line(5,0)(320,0)
  \LongArrow(10,-10)(10,100)
  \Text(0,105)[b]{mass}
  \Text(0,0)[r]{$0$}
  \Line(8,20)(12,20)    \Text(6,20)[r]{$1/R$}
  \Line(8,40)(12,40)    \Text(6,40)[r]{$2/R$}
  \Line(8,60)(12,60)    \Text(6,60)[r]{$3/R$}
  \Line(8,80)(12,80)    \Text(6,80)[r]{$4/R$}
  \Text(60,115)[b]{$(+,+)$}
  \Text(60,100)[b]{$q, h, A_\mu$}
  \Line(40,0)(80,0)      \Vertex(60,0){3}
  \Line(40,40)(80,40)    \Vertex(60,40){3}
  \Line(40,80)(80,80)    \Vertex(60,80){3}
  \Text(130,115)[b]{$(+,-)$}
  \Text(130,100)[b]{$\tilde{q}, \tilde{h}, \lambda$}
  \Line(110,20)(150,20)    \Vertex(130,20){3}
  \Line(110,60)(150,60)    \Vertex(130,60){3}
  \Text(200,115)[b]{$(-,+)$}
  \Text(200,100)[b]{$\tilde{q}^c, \tilde{h}^c, \lambda^c$}
  \Line(180,20)(220,20)    \Vertex(200,20){3}
  \Line(180,60)(220,60)    \Vertex(200,60){3}
  \Text(270,115)[b]{$(-,-)$}
  \Text(270,100)[b]{$q^c, h^c, \sigma$}
  \Line(250,40)(290,40)    \Vertex(270,40){3}
  \Line(250,80)(290,80)    \Vertex(270,80){3}
\end{picture}
\caption{Tree-level KK mass spectrum of the matter, Higgs and gauge 
multiplets in $S^1/(Z_2 \times Z_2')$ compactification.}
\label{fig:S1Z2Z2}
\end{center}
\end{figure}
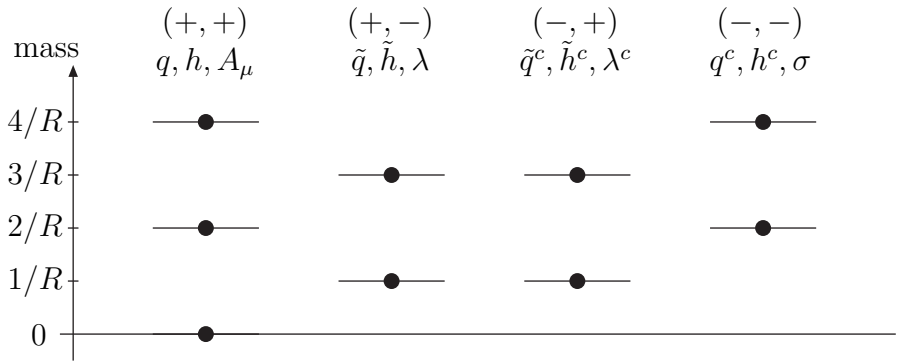

The KK spectrum of the model is given in Fig.~\ref{fig:S1Z2Z2}.
The zero-mode sector consists of the standard-model fields and, at the 
first excitation level, we have two superparticles for each 
standard-model particle reflecting underlying 5D SUSY structure.
In particular, there are two superpartners (two Higgsinos) for the 
Higgs boson, which are massive forming a Dirac mass term, so that 
the theory is anomaly free.  Note that we have evaded usual argument 
leading to two Higgs doublets simply because there is no energy interval 
where the 4D ${\cal N}=1$ supersymmetric description is appropriate 
in this theory.

How about interactions?  
The Yukawa couplings can be introduced on the fixed points of the 
orbifold.  We require that they preserve the symmetries which remain 
unbroken after the orbifold reflection about that fixed point.
For instance, $Z_2$ orbifolding preserves 4D ${\cal N}=1$ SUSY, so that 
the interactions placed on the $y=0$ fixed point must preserve this 
symmetry.  Choosing the hypercharge for $h$ appropriately, we can 
place the supersymmetric Yukawa interactions $[ Q U H ]_{\theta^2}$ 
on the $(3+1)$-dimensional hypersurface at $y=0$, which contain 
the up-type Yukawa couplings $quh$ of the standard model.  Here, 
$Q$, $U$ and $H$ represent chiral superfields under the 4D ${\cal N}=1$ 
SUSY that remains unbroken after the $Z_2$ orbifolding.  Specifically,
\begin{eqnarray}
  Q &=& \tilde{q} + \theta\, q + \cdots, 
\\
  H &=& h + \theta\, \tilde{h} + \cdots,
\label{eq:cs-Z2}
\end{eqnarray}
and they are represented by solid lines in Fig.~\ref{fig:transf}.
Similarly, we can also introduce interactions at the $y=\pi R/2$ fixed 
point.  However, an important point here is that the remaining 4D 
${\cal N}=1$ SUSY after the $Z_2'$ orbifolding is different from that 
after the $Z_2$.  Thus, we must use {\it different} chiral superfields 
to write down the interactions at $y=\pi R/2$.  They are denoted by 
dashed lines in Fig.~\ref{fig:transf}, and given, for example, by
\begin{eqnarray}
  Q' &=& \tilde{q}^{c\dagger} + \theta'\, q + \cdots, 
\\
  H^{c\prime} &=& h^\dagger + \theta'\, \tilde{h}^c + \cdots.
\label{eq:cs-Z2'}
\end{eqnarray}
Then, the interactions we can write down at $y=\pi R/2$ are 
$[ Q' D' H^{c\prime} ]_{\theta^{\prime 2}}$, which contain 
the down-type Yukawa couplings $qdh^\dagger$ (and similarly for the 
charged leptons).  Therefore, the interactions of the model are also 
those of the standard model, but the tree-level Higgs potential takes 
special form given in Eq.~(\ref{eq:Higgs-tree}) since it only comes 
from the $D$-term potential of the gauge interactions.

\subsection{Finite radiative electroweak symmetry breaking}

Having established the tree-level structure of the model, we are now 
at the position of discussing radiative corrections.
As usual, the dominant radiative correction to the Higgs potential 
comes from the loops of the top quarks and squarks through the Yukawa 
couplings.  Here, however, all the KK towers are circulating the loops, 
and it makes the radiative correction finite.
The one-loop effective potential $V_{h, {\rm 1-loop}}$ is given by 
\cite{Barbieri:2001vh}
\begin{equation}
  V_{h, {\rm 1-loop}} = {18 \over \pi^6 R^4} 
    \sum_{k=0}^\infty {\cos[2(2k+1) 
    \arctan \left( \pi y_t R h / 2 \right)] \over (2k+1)^5},
\label{eq:Higgs-1loop}
\end{equation}
where $y_t$ is related to the top-quark mass by $m_t = (2/\pi R) 
\arctan ( \pi y_t R \left\langle h \right\rangle / 2 )$.
Note that the momentum integral leading to the above expression is 
completely dominated by $1/R$ scale, indicating that the physics of 
EWSB lies at the compactification scale.  Furthermore, the summation 
of KK towers can also be cut off at a first few levels without affecting 
the result much.  An important point here is that this cut-off procedure 
(regularization) must preserve correct symmetries of the theory 
\cite{Barbieri:2001vh}.  Simply cutting off the summation at a finite 
level does not work, since it does not respect SUSY \cite{Ghilencea:2001ug}.
Appropriate regularizations have recently been discussed in 
Refs.~\cite{Delgado:2001ex}.

The above points are elegantly understood by calculating the Higgs mass 
in the 5D mixed position-momentum space \cite{Arkani-Hamed:2001mi}.
In this calculation, we work in the momentum space for usual 4D and 
the position space for the fifth dimension.  We find that the diagrams 
in which the internal loops can continuously shrink to a point are 
cancelled between the bosonic and fermionic loops, and the only diagrams 
in which the internal propagators non-trivially wind round the extra 
dimension contribute to the Higgs mass.  Then, since propagators are 
given by $\sim \exp(-p|y|)$, where $p$ is the 4-momentum, the 
contributions from the modes with $p \gsim (\pi R)^{-1}$ are 
exponentially suppressed.  (This precisely corresponds to the 
point-splitting regularization of a distance $\pi R$.)
This shows that EWSB is caused by the physics at the compactification 
scale (boundary conditions) and exponentially insensitive to the physics 
at ultraviolet scales.  An intuitive understanding of the result is 
the following.  Since SUSY is broken by the boundary conditions, 
the Higgs boson could know this non-local breaking only when the 
internal top/stop propagators go around the circle.  Then, it is clear 
that the radiative correction to the Higgs mass is finite, since the 
internal loops cannot shrink to a point where ultraviolet divergences 
would arise.

There is another way of understanding the finiteness of the Higgs mass.
Remember that there are unbroken 4D ${\cal N}=1$ SUSYs both at the $y=0$ 
and $y=\pi R/2$ fixed points, although they are different subgroups of 
the original 5D SUSY.  Likewise, we can also show that there are some 
residual SUSYs in {\it any} point in the extra dimension.
(Of course, these SUSYs are different at different points in the 
 extra dimension, so that there is no {\it global} SUSY after 
 integrating out the extra dimension.)
This means that we cannot write down a local operator which gives 
the Higgs mass.  Therefore, if we calculate the Higgs boson mass, it 
must be finite since there is no counterterm for it.
Here, by finite, we mean that there is no {\it intrinsic} divergence 
in the Higgs mass.  Suppose that we calculate higher loop diagrams.
Then, there are, of course, sub-divergences which give divergent 
contributions to the Higgs mass.  However, they are all subtracted by 
supersymmetric counterterms.  In other words, after renormalizing all 
supersymmetric quantities such as gauge and Yukawa couplings, the 
expression of the Higgs mass is completely finite at all orders.

One final comment on the finiteness of radiative corrections is in order.
From the above arguments, it should be clear that the finiteness of the 
Higgs mass relies on the fact that local SUSYs prevent to write 
the counterterm for the Higgs mass.  This means that the quantities 
associated with SUSY breaking, {\it i.e.} the quantities which are 
vanishing in the SUSY limit, are all calculable in the present model.  
On the other hand, the quantities which can take nonvanishing values 
in the SUSY limit, such as the $\rho$ parameter, are generically divergent 
and uncalculable \cite{Barbieri:2001vh}.

\subsection{Phenomenology}

We can now calculate the compactification radius $R$ and the physical 
Higgs boson mass $m_{{\rm phys},h}$ by minimizing the Higgs potential 
obtained from Eq.~(\ref{eq:Higgs-tree}) and Eq.~(\ref{eq:Higgs-1loop}).
They are given by
\begin{eqnarray}
  \frac{1}{R} &=& (1 + z) (352 \pm 20) ~{\rm GeV},
\\
  m_{{\rm phys},h} &=& 127 \pm 8 ~{\rm GeV},
\label{eq:pred}
\end{eqnarray}
where $z$ ($|z| \lsim 0.2$) parameterizes unknown effects from 
ultraviolet physics and the errors come from 2-loop contributions 
etc \cite{Barbieri:2001vh}.  This value of $R$ is phenomenologically 
acceptable because there is no strong constraint coming from the 
direct production of KK gauge bosons or generation of four fermion 
operators in the models where the quarks and leptons propagate 
in the bulk \cite{Barbieri:2001vh,Appelquist:2000nn}.

The masses for the superparticles are shifted from $1/R$ after EWSB.
The largest effect appears on the top squark sector.  Among the four 
top squarks (note that the number of superparticles are doubled compared 
with MSSM), two become heavier and two lighter after EWSB.  The latter
provides the lightest superparticle of mass 
\begin{eqnarray}
  m_{\tilde{t}_-} &=& 197 \pm 20 ~{\rm GeV}.
\label{eq:stop}
\end{eqnarray}
Since these particles are stable, they produce characteristic signals 
in hadron colliders such as highly ionizing tracks and intermittent 
highly ionizing tracks \cite{Barbieri:2001vh}.  Incidentally, the 
present limit on such particles is about 150 GeV from CDF 
\cite{Connolly:1999dv}.

\section{Conclusions}
\label{sec:concl}

If there is a TeV-sized extra dimension, we can construct 
supersymmetric theories with (i) no hidden sector (ii) finite 
radiative Higgs mass (iii) one Higgs doublet.
In a minimal such model based on the $S^1/(Z_2 \times Z_2')$ orbifold, 
the compactification radius and the physical Higgs boson mass are 
determined as $1/R \simeq 350$ GeV and $m_{{\rm phys},h} \simeq 127$ GeV.
The lightest superparticle is a top squark of mass 
$m_{\tilde{t}_-} \simeq 197$ GeV, which may be seen at future linear 
collider experiments.

\section*{Acknowledgments}
The author would like to thank N.~Arkani-Hamed, R.~Barbieri, L.~Hall, 
D.~Smith and N.~Weiner for fruitful collaborations and valuable 
discussions on the subject.  This work was supported by the Miller 
Institute for Basic Research in Science.

\end{document}